\documentclass[aps,prl,showpacs,twocolumn]{revtex4}
\usepackage{graphicx}
\usepackage[T1]{fontenc} 
\usepackage{epsfig,latexsym}
\usepackage{color}
\usepackage[latin1]{inputenc}
\usepackage{amsmath}
\usepackage{amssymb}
\usepackage{amsfonts}
\usepackage{indentfirst}
\usepackage{ae}
\usepackage{float}

\begin{document}
\title{Lorentz Covariant Quantum 4-Potential and Orbital Angular Momentum for the Transverse Confinement of Matter Waves}
\author{R. Ducharme$^{1}$ and I. G. da Paz$^{2}$}

\affiliation{$^{1}$ 2112 Oakmeadow Pl., Bedford, TX 76021, USA}

\affiliation{$^2$ Departamento de F\'{\i}sica, Universidade Federal
do Piau\'{\i}, Campus Ministro Petr\^{o}nio Portela, CEP 64049-550,
Teresina, PI, Brazil}

\begin{abstract}
In two recent papers exact Hermite-Gaussian solutions to relativistic 
wave equations have been obtained for both electromagnetic and particle 
beams that include Gouy phase. The solutions for particle beams correspond 
to those of the Schr\"{o}dinger equation in the non-relativistic limit. Here,
distinct canonical and kinetic 4-momentum operators will be defined for 
quantum particles in matter wave beams. The kinetic momentum is
equal to the canonical momentum minus the fluctuating terms resulting
from the transverse localization of the beam. Three results are obtained. 
First, the total energy of a particle for each beam mode is calculated. 
Second, the localization terms couple into the canonical 4-momentum of 
the beam particles as a Lorentz covariant quantum 4-potential originating 
at the waist. The quantum 4-potential plays an analogous role in relativistic 
Hamiltonian quantum mechanics to the Bohm potential in the non-relativistic 
quantum Hamilton-Jacobi equation. Third, the orbital angular momentum (OAM)
operator must be defined in terms of canonical momentum operators.  It
is further shown that kinetic 4-momentum does not contribute to OAM
indicating that OAM can therefore be regarded as a pure manifestation
of quantum 4-potential.
\end{abstract}


\maketitle

\section{I. Introduction}
Experiment shows that beams of particles still behave like beams
even if only one particle is traveling through the apparatus at a
time \cite{MMP}. The converse of this argument is that isolated
particles can behave like beams. Specifically, it is understood that
the wavefunction $\Psi$ for the particle must take account of a full
compliment of wave beam features such as mode numbers
\cite{siegman}, Gouy phase \cite{gouy1, gouy2} and orbital angular
momentum \cite{Allen}.

The purpose of this paper is to explain the localizing affect of 
transverse confinement on a beam particle using a quantum 4-potential.  
The concept of a quantum 4-potential as it is introduced here is similar
to the Bohm potential \cite{Bohm} in the the sense that it is a
concept extracted from $\Psi$ rather than a representation of a field
separate from $\Psi$. The point of departure is that the quantum
4-potential couples into each individual component of the 4-momentum
operators for the particle whereas the scalar Bohm potential is a
component in a Hamilton-Jacobi equation belonging to an alternate
formulation of quantum mechanics. The quantum 4-potential
therefore requires a distinction to be made between the canonical
$\hat{p}_{\mu} (\mu=1,2,3,4)$ and kinetic $\hat{P}_{\mu}$ 4-momentum
for the particle. The canonical (total) 4-momentum is the sum of the
kinetic 4-momentum and the quantum 4-potential term.

Free particles have no quantum potential but localized particles do
have it. The signature of a quantum potential is therefore the appearance 
a term in a quantum mechanical equation that generates localization and
has no association to external source. It will be shown that the form of 
this term depends on the specific formulation of quantum mechanics under 
consideration but that all the variants interrelate and have two distinct 
common properties; they vanish in the free particle limit and have null 
expectation values.

External devices are responsible for collimating and focusing 
the particles in a beam. Once a particle has passed through these 
devices, it remains localized but is no longer confined. Our
solutions describe the localized state of the particles but not
the passage of the particle through the devices responsible for
confining the beam. 

The basic structure of a wave beam can be understood using the
Heisenberg uncertainty principle \cite{feng2001} that states
uncertainty in momentum is inversely proportional to uncertainty in
position. In a continuous wave beam there is no localization of the
particle along the axis of the beam meaning that each particle can
be assumed to have a precise axial momentum and therefore a precise
axial velocity $v_3$. The uncertainty in the position of the
particle along the transverse axis is smallest at the beam waist. It
is therefore the size of the waist that determines the uncertainty
in the transverse momentum of the particle. The presence of
transverse momentum explains the fact beams spread. It also accounts
for the existence of orbital angular momentum in beams.

Linear wave equations have both plane-wave and localized solutions
\cite{APK} often called wave packets \cite{MVA}. The wave packet is
smallest at the time of an event that localizes the particle then
continuously grows in size afterwards. One distinguishing
characteristic of plane-wave and localized wave functions is the
number of 4-position dependencies in them. Plane-waves are local
functions that only depend on the position $x_i (i=1,2,3)$ of the
particle at time $t$. By contrast, localized wave solutions are
bilocal functions since they must depend on both the current
4-position of the particle as well as the 4-position $X_\mu = (X_i,
cT)$ of the preceding confinement event where and when the size of
the wave packet was at a minimum. It is the bilocal nature of wave
packets that permits the probability density of finding free
particles to have spatial extension as well as a 4-position. It is
also the dependence of $\Psi$ on $X_\mu$ as well as $x_\mu$ that
will enable us to define distinct kinetic $P_\mu$ and canonical $p_\mu$
4-momentum vectors.

Bateman-Hillion functions \cite{PH, BS} are exact localized
solutions of relativistic wave equations that trace back to early
work of Bateman on conformal transformations \cite{HB}. In two
recent papers, exact Bateman-Hillion solutions were obtained for the
Hermite-Gaussian modes of both electromagnetic \cite{RD1} and
quantum particle \cite{RD2} beams. These are detailed solutions for
particle beams that include Gouy phase \cite{PNP, Paz2,cond}. The
paraxial wave equation \cite{siegman} for electromagnetic beams and
the Schr\"{o}dinger equation for non-relativistic particle beams
have both been demonstrated as limiting cases of the Bateman-Hillion
method.

One method of obtaining Bateman-Hillion solutions to a wave equation
is to start from an ansatz. In the case of the Klein-Gordon equation
the ansatz eliminates the second order time derivative reducing the
wave equation to a parabolic form.  This resolves problems of
negative energies and negative probability densities that afflict
the unconstrained Klein-Gordon equation. It will be further shown in
this paper that the probability density of finding a particle in a
Bateman-Hillion beam is just $|\Psi|^2$ similar to the
Schr\"{o}dinger equation except that the probability density for
Bateman-Hillion solutions is also form preserving under Lorentz
transformations.

In this paper a transformation will be made to the Bateman-Hillion
solutions of the Klein-Gordon equation for particle beams to account
for an earlier finding \cite{RD2} that the components of the
4-momentum of the particles must have a shift in them related to the
complex shift in the 4-position coordinates needed for the accurate
description of any wave beam. This will be shown to facilitate a
calculation for the total energy of each particle in terms of the
rest mass of the particle, the kinetic energy of the propagation of
the particle along the axis of the beam and the kinetic energy
locked up in the transverse mass flows. Results will be presented
for both Hermite-Gaussian and Laguerre-Gaussian beams.
Laguerre-Gaussian beams are useful to describe the orbital angular
momentum states of the particle.

After the seminal paper by Bliokh at al introducing vortex beams
carrying OAM for free quantum electrons \cite{Bliokh1} several
experimental \cite{Uchida} and theoretical \cite{Verbeeck, BDN} results
were obtained. Properties of the interaction of OAM with an electric
field such as OAM Hall effect was studied in the non relativistic
context \cite{Bliokh1}. Further the interaction of OAM with a
magnetic field was also studied in the non relativistic context
\cite{Bliokh2}. More recently the effect of the interaction of
relativistic electron vortex beam with a laser field was studied
showing that the beam center is shifted and that the shift in the
paraxial beams is larger than that in the nonparaxial beams
\cite{AGH, Pratul}. The results that we are obtaining in this paper could
be useful to explore the relativistic effects in the properties such
as OAM Hall and Zeeman effect resulting, respectively, of the
interaction of relativistic scalar (without spin) electron vortex
beam with an electric and magnetic field. Further we can similarly
solve the Dirac equation to include the effects of the interaction
of spin angular momentum (SAM) with a magnetic field.

It will be shown in this paper that the Schr\"{o}dinger and 
Klein-Gordon equations give the same orbital angular momentum
for each scalar mode of a Laguerre-Gaussian beam. To find relativistic 
corrections to orbital angular momentum it is therefore necessary to 
investigate solutions that mix multiple modes. For example, in the 
case of Bessel beam solutions to the Dirac equation it has been found 
\cite{BDN} the corrective amplitude coefficients take the form 
$a = \sqrt{1-E_0/E)} \sin\theta_0$ where $E$ denotes the energy of
each particle, $E_0$ is the rest energy and $\theta_0$ is the polar
angle indicating the divergence of the beam. This results in a 
relativistic  correction $sa^2$  to the total angular momentum of each 
particle with spin $s$. The correction clearly vanishes in both the 
non-relativistic $(E \rightarrow E_0)$ and paraxial 
$(\theta_0 \rightarrow 0)$ limits but can otherwise affect the 
energies of beam particles in external electric and magnetic fields. 
Another source for relativistic corrections that may affect OAM is the 
repulsion between charged particles. This can be a stronger effect 
than the spin-orbit interaction that could be studied using either 
the Klein-Gordon or Dirac equations. The repulsion between charged 
particles is also known to have a greater affect on the beam for lower 
energy particles.

The fact $\Psi(x_i, t, X_i, T)$ depends on two 4-position vectors
requires the introduction of a constraint condition \cite{AK, CA} to
eliminate one of the independent time coordinates in the calculation
of the physical properties for the beam. As in an earlier paper
\cite{RD2} the solution to be applied here is to use Dirac delta
function notation to impose a relationship $\xi_3 - v_3 \tau = 0$
between the relative position $\xi_i= x_i-X_i$ and relative time
$\tau = t - T$. This relates back to the idea that particles in
continuous wave beams can be assigned a precise axial velocity
$v_3$.

In Sec. II, we use the Bateman-Hillion ansatz to solve the
Klein-Gordon equation for a particle that passes through a beam
waist . In Sec. III, we determine the Lorentz invariant probability
density of finding a particle in a Bateman-Hillion beam. In Sec. IV,
we calculate the kinetic 4-momentum in terms of the canonical
4-momentum and the localization terms. In Sec. V, we calculate the
quantum 4-potential. In Sec. VI, we conclude our results in a
summary.

\section{II. Bateman-Hillion Beams}
Consider a beam of particles each having a rest mass $m_0$, a
4-position $x_\mu = (x_i, ct)$ and a 4-momentum $p_\mu = (p_i, E /
c)$. Let us assume each particle passes through a beam waist
with a position $X_i$ at the time $T$. The
Klein-Gordon equation for the wave function $\Psi(x_i, t, X_i, T)$
representing each of the particles in Minkowski space can be
expressed as
\begin{equation} \label{eq: KG}
\hat{p}_\mu \hat{p}^\mu \Psi = \frac{1}{c^2}(\hat{E}^2 - c^2\hat{p}_i^2)\Psi =
m_0^2c^2\Psi,
\end{equation}
where
\begin{equation} \label{eq: canonical_momentum_operators}
\hat{p_i} = \frac{\hbar}{\imath} \frac{\partial}{\partial x_i},
\quad \hat{E} = -\frac{\hbar}{\imath} \frac{\partial}{\partial t},
\end{equation}
are the canonical 4-momentum operators, $\hbar$ is Planck's constant
divided by $2\pi$ and $c$ is the velocity of light.

One approach to solving eq. (\ref{eq: KG}) for a beam is to use a Bateman 
inspired ansatz. In an earlier paper \cite{RD2}, the following trial form 
was taken as the starting point for the derivation of the positive-energy 
Hermite-Gaussian beam solutions
\begin{equation} \label{eq: bateman_solution_old}
\Psi_{mn}^{O} = \Phi_{mn}(\xi_1, \xi_2, \xi_3+c\tau) \exp[ \imath(k_3 x_3 -k_4 ct)],
\end{equation}
where
\begin{equation} \label{eq: relative_coordinates}
\xi_i = x_i - X_i, \quad \tau = t -  T,
\end{equation}
gives the position of each point $x_\mu$ relative to the 4-position
of the beam waist, $k_\mu = (0, 0, k_3, k_4)$ is the wave
vector and $\Phi_{mn}$ are scalar functions. The positive integers
$m$ and $n$ indicate the mode of the beam.

A curious feature of eq. (\ref{eq: bateman_solution_old}) derived in \cite{RD2} is that
it leads to the following expression for the particle current in a Gaussian beam
\begin{equation} \label{eq: four_momentum_old}
\langle \Psi_{00}^{O} | \hat{j}_\mu | \Psi_{00}^{O} \rangle = \frac{\hbar}{m_0}(k_\mu - \kappa_\mu^{00}),
\end{equation}
where
\begin{equation} \label{eq: particle_current_def}
\Psi^* \hat{j}_\mu \Psi = \frac{1}{2m_0} (\Psi^* \hat{p}_\mu \Psi -
\Psi \hat{p}_\mu \Psi^*),
\end{equation}
and $\kappa_\mu^{mn} = (0, 0, \kappa^{mn}, -\kappa^{mn})$. Here, the axial parameter $\kappa^{mn}$ takes the form
\begin{equation} \label{eq: kappa_defnition_00}
\kappa^{00} = \frac{1}{(k_3 + k_4)w_0^2},
\end{equation}
where $w_0$ is the radius of the beam at the waist.

Eq. (\ref{eq: four_momentum_old}) suggests that $k_\mu$ is related to the
expectation value of the axial current for a particle in a beam.
In seeking an intuitive definition for $k_\mu$ we shall now make use of
the unitary transformation
\begin{equation} \label{eq: unitary_transformation}
\Psi_{mn} = \Psi_{mn}^O \exp \left[ \imath \kappa^{mn} (x_3 + ct)
\right],
\end{equation}
where
\begin{equation} \label{eq: kappa_defnition}
\kappa^{mn} = \frac{N^{mn}}{(k_3 + k_4)w_0^2},
\end{equation}
and $N^{mn}$ is a constant.  The general form of $N^{mn}$ is to be
determined but it can be seen from comparison of eqs.
(\ref{eq: kappa_defnition_00}) and (\ref{eq: kappa_defnition}) 
that $N^{00} = 1$. It is also readily verified that eq. (\ref{eq: unitary_transformation}) 
is form invariant under the Lorentz transformation equations:
\begin{equation} \label{eq: lorentz_x}
x_3^\prime = (x_3 - v_3 \tau) \gamma, \quad \tau^\prime = (\tau - \frac{v_3}{c^2} x_3) \gamma
\end{equation}
\begin{equation} \label{eq: lorentz_k}
k_3^\prime = (k_3 - \frac{v_3}{c} k_4) \gamma, \quad k_4^\prime = (k_4 - \frac{v_3}{c} k_3) \gamma
\end{equation}
where $\gamma = 1 / \sqrt{1-v_3^2 /c^2}$. Applying the
transformation (\ref{eq: unitary_transformation}) to eq. (\ref{eq:
bateman_solution_old}) gives
\begin{multline} \label{eq: bateman_solution}
\Psi_{mn} = \Phi_{mn}(\xi_1, \xi_2, \xi_3+c\tau) \\
\times \exp\left[ \imath (k_3+\kappa^{mn}) x_3 - \imath c(k_4-\kappa^{mn})t
\right],
\end{multline}
equivalent to making the replacements $k_3 \rightarrow k_3 + \kappa^{mn}$
and $k_4 \rightarrow k_4 - \kappa^{mn}$. These replacements can be used, in turn to reduce
eq. (\ref{eq: four_momentum_old}) to the simplified to the form
\begin{equation} \label{eq: four_momentum_refined}
\langle \Psi_{00} | \hat{j}_\mu | \Psi_{00} \rangle = \frac{\hbar}{m_0} \left(
0, 0, k_3, k_4 \right),
\end{equation}
where it can be seen $\kappa^{00}$ has been eliminated. One important goal of this
paper will be to show that there exists $N^{mn}$ such that the condition
\begin{equation} \label{eq: four_momentum_expected}
\langle \Psi_{mn} | \hat{j}_\mu | \Psi_{mn} \rangle = \frac{\hbar}{m_0} \left(
0, 0, k_3, k_4 \right),
\end{equation}
is satisfied. If this hypothesis is true, it implies $\frac{\hbar}{m_0} k_\mu$ 
can be interpreted as the expectation value for the particle current in a relativistic 
beam thus giving a clear physical meaning to $k_\mu$.

Inserting eq. (\ref{eq: bateman_solution}) into the Klein-Gordon
equation (\ref{eq: KG}) gives
\begin{multline} \label{eq: KG_bateman}
\frac{\partial^2 \Phi_{mn}}{\partial x_1^2} +
\frac{\partial^2 \Phi_{mn}}{\partial x_2^2} +  2 \imath \left( k_3 + \kappa^{mn} \right)
\frac{\partial \Phi_{mn}}{\partial x_3} \\
+  \frac{2 \imath}{c} \left( k_4 - \kappa^{mn}
\right)\frac{\partial \Phi_{mn}}{\partial t}  = 0,
\end{multline}
where
\begin{equation} \label{eq: dispersion_relation}
k_4^2 =  k_3^2 + 2\kappa^{mn} \left( k_3+k_4 \right) + \frac{m_0^2 c^2}{\hbar^2}.
\end{equation}
It can be seen the unitary transformation (\ref{eq: unitary_transformation}) has introduced the term
\begin{equation} \label{eq: mystery_term}
K_T^{mn} = 2 \kappa^{mn} \left( k_3 + k_4 \right),
\end{equation}
into this dispersion relationship. The physical interpretation of
$K_T^{mn}$ will be discussed later once the relativistic energy formula
for each particle in the beam has been derived.

It is instructive to observe that
\begin{equation} \label{eq: operator_equivalence_1}
\frac{\partial}{\partial x_3} \Phi_{mn} = \frac{1}{c}
\frac{\partial}{\partial t} \Phi_{mn},
\end{equation}
and equivalently
\begin{equation} \label{eq: operator_equivalence_2}
\frac{\partial}{\partial x_3} |\Psi_{mn}|^2 = \frac{1}{c}
\frac{\partial}{\partial t} |\Psi_{mn}|^2,
\end{equation}
owing the fact $\Phi_{mn}$ only depends on $\xi_3$ and $\tau$ in the linear combination $\xi_3+\tau$.
Eqs. (\ref{eq: KG_bateman}) and (\ref{eq: operator_equivalence_1}) can now be combined to obtain the operator relationships
\begin{equation} \label{eq: operator_relationship_3}
\hat{p}_3 \Phi_{mn} = -\hat{p}_4 \Phi_{mn} = -\frac{\hat{p}_1^2 + \hat{p}_2^2}{2\hbar(k_3+k_4)} \Phi_{mn},
\end{equation}
These results will prove useful later.

Equation (\ref{eq: KG_bateman}) can be solved analogously to the paraxial
equation \cite{siegman} to give
\begin{multline} \label{eq: hermite_gauss_solution}
\Phi_{mn} = \frac{C_{mn}^{HG} w_0}{w}H_m\left(
\frac{\sqrt{2}\xi_1}{w}\right) H_n\left(
\frac{\sqrt{2}\xi_2}{w}\right)\times\\ \exp \left[ \frac{ \imath
2b (\xi_1^2+\xi_2^2)}{w_0^2(\xi_3+c\tau - \imath 2b)} - \imath
g_{mn} \right],
\end{multline}
where $H_m$ and $H_n$ are Hermite polynomials,
\begin{equation} \label{eq: b_term}
b = \frac{w_0^2}{4} \left( k_3 + k_4\right),
\end{equation}
\begin{equation} \label{eq: spot_radius}
w(\xi_3,\tau) = w_0\sqrt{1+\left( \frac{\xi_3+c\tau}{2b} \right)^2},
\end{equation}
is the beam radius such that $w_0=w(0,0)$ and
\begin{equation} \label{eq: gouy_phase_mn}
g_{mn}(\xi_3,\tau) = (1+m+n)\arctan \left( \frac{\xi_3+c\tau}{2b} \right),
\end{equation}
is the Gouy phase of a relativistic quantum particle.

It is notable that the Klein-Gordon equation (\ref{eq: KG}) can also be usefully solved in
cylindrical coordinates starting from the expression
\begin{multline} \label{eq: bateman_solution_cyl}
\Psi_{lp} = \Phi_{lp}(\xi_\rho, \xi_\phi, \xi_3+c\tau) \\
\times \exp\left[ \imath (k_3+\kappa^{lp}) x_3 - \imath c(k_4-\kappa^{lp})t \right]
\end{multline}
equivalent to eq. (\ref{eq: bateman_solution}) where $\xi_\rho = \sqrt{\xi_1^2+\xi_2^2}$ and $\xi_\phi = $atan2$(\xi_2, \xi_1)$. This gives
\begin{multline} \label{eq: laguerre_gauss_solution}
\Phi_{lp} = \frac{C_{lp}^{LG} w_0}{w}\left(
\frac{\sqrt{2}\xi_\rho}{w}\right)^{|l|} L_p^{|l|}\left(
\frac{2 \xi_\rho^2}{w^2}\right) \times\\ \exp \left[ \frac{ \imath
2b \xi_\rho^2}{w_0^2(\xi_3+c\tau - \imath 2b)} +\imath l \xi_{\phi}- \imath
g_{lp} \right],
\end{multline}
where $L_p^{|l|}$ are the generalized Laguerre polynomials and
\begin{equation} \label{eq: gouy_phase_lp}
g_{lp}(\xi_3,\tau) = (1+|l|+2p)\arctan \left( \frac{\xi_3+c\tau}{2b} \right),
\end{equation}
is the Gouy phase in terms of the radial Laguerre index $p$ and the azimuthal index $l$ that may be positive or negative.

The operator for the axial component of canonical OAM can
be expressed as
\begin{equation} \label{eq: OAM_operator}
\hat{L}_3 = \xi_\rho \times \hat{p}_{\phi} = \frac{\hbar}{\imath} \frac{\partial}{\partial \xi_\phi}
\end{equation}
The Laguerre-Gaussian beam functions (\ref{eq: bateman_solution_cyl}) can
thus be seen to give
\begin{equation} \label{eq: angular_momentum}
\hat{L}_3 \Psi_{lp} = l \hbar \Psi_{lp},
\end{equation}
showing $L_{3}=l\hbar$ are the possible eigenvalues of OAM for a
Laguerre-Gaussian beam.

\section{III. Probabilistic Interpretation}
In this section, the correspondence between the particle current
(\ref{eq: particle_current_def}) for Bateman-Hillion beams and that
of the Schr\"{o}dinger equation for particle beams will be
investigated as means of determining the probability density of
finding a particle in a Bateman-Hillion beam. As a starting point it
will be useful to evaluate each component of the Bateman-Hillion
particle current
\begin{equation}
j_\mu^{mn}=\Psi_{mn}^* \hat{j}_\mu \Psi_{mn}.
\end{equation}
This leads to
\begin{equation} \label{eq: paricle_current_1}
j_1^{mn} = \frac{4b(\xi_3+c \tau) \xi_1}{w_0^2[(\xi_3+c \tau)^2 +4b^2]}\frac{\hbar}{m_0} |\Psi_{mn}|^2,
\end{equation}
\begin{equation} \label{eq: paricle_current_2}
j_2^{mn} = \frac{4b(\xi_3+c \tau) \xi_2}{w_0^2[(\xi_3+c \tau)^2 +4b^2]}\frac{\hbar}{m_0} |\Psi_{mn}|^2,
\end{equation}
\begin{multline}\label{eq: paricle_current_3}
j_3^{mn} = \left[ k_3 + \kappa^{mn}
- \frac{2b(1+m+n)}{(\xi_3+c \tau)^2+4b^2} \right] \frac{\hbar}{m_0}  |\Psi_{mn}|^2\\
 - \frac{2b(\xi_1^2+\xi_2^2)[(\xi_3+c \tau)^2-4b^2]}{w_0^2[(\xi_3+c \tau)^2+4b^2]^2} \frac{\hbar}{m_0} |\Psi_{mn}|^2,
\end{multline}
\begin{multline}\label{eq: paricle_current_4}
j_4^{mn} = \left[ k_4 - \kappa^{mn} + \frac{2b(1+m+n)}{(\xi_3+c \tau)^2+4b^2} \right] \frac{\hbar}{m_0} |\Psi_{mn}|^2 \\
+ \frac{2b(\xi_1^2+\xi_2^2)[(\xi_3+c \tau)^2-4b^2]}{w_0^2[(\xi_3+c \tau)^2+4b^2]^2}\frac{\hbar}{m_0} |\Psi_{mn}|^2,
\end{multline}
where
\begin{multline} \label{eq: prob_dens__KG}
|\Psi_{mn}|^2 = \left(\frac{C_{mn}^{HG} w_0}{w} \right)^2 H_m^2\left(
\frac{\sqrt{2}\xi_1}{w}\right) H_n^2 \left(
\frac{\sqrt{2}\xi_2}{w}\right)\\
\times \exp \left[ -\frac{8 b^2 (\xi_1^2+\xi_2^2)}{w_0^2[(\xi_3+c
\tau)^2+4b^2]} \right].
\end{multline}

The continuity equation for the Klein-Gordon equation (\ref{eq: KG}) is
\begin{equation} \label{eq: continuity_KG}
\frac{\partial j_1}{\partial x_1} + \frac{\partial j_2}{\partial
x_2} + \frac{\partial j_3}{\partial x_3} +  \frac{1}{c}
\frac{\partial j_4}{\partial t}   = 0.
\end{equation}
Eqs. (\ref{eq: paricle_current_3}) and (\ref{eq: paricle_current_4}) enable this expression to be rewritten in the form
\begin{equation} \label{eq: continuity_BH}
\frac{\partial j_1}{\partial x_1} +
\frac{\partial j_2}{\partial x_2} +
\frac{1}{m_0}\left( k_3 \frac{\partial}{\partial x_3} +  k_4 \frac{\partial}{\partial t} \right) |\Psi_{mn}|^2  = 0,
\end{equation}
or equivalently
\begin{equation} \label{eq: continuity_BH_time_form}
\frac{\partial j_1}{\partial x_1} +
\frac{\partial j_2}{\partial x_2} +
\frac{1}{m_0}\left( k_3 +  k_4  \right) \frac{\partial}{\partial t} |\Psi_{mn}|^2  = 0,
\end{equation}
having used eq. (\ref{eq: operator_equivalence_2}). This result reduces to the simplified expression
\begin{equation} \label{eq: continuity_Schrodinger}
\frac{\partial j_1}{\partial x_1} +
\frac{\partial j_2}{\partial x_2} +
\frac{\partial}{\partial t} |\Psi_{mn}^S|^2  = 0,
\end{equation}
in the non-relativistic limit where $k_3 \ll k_4$ and $m_0c^2 \simeq c \hbar k_4$.

In an earlier paper \cite{RD2} it was shown that eqs. (\ref{eq: KG}) and (\ref{eq: bateman_solution_old}) reduce to the Schr\"{o}dinger equation
\begin{equation} \label{eq: schrodinger_wave_equation}
\frac{\partial^2 \Psi_{mn}^S}{\partial x_1^2} + \frac{\partial^2
\Psi_{mn}^S}{\partial x_2^2} + \frac{\partial^2
\Psi_{mn}^S}{\partial x_3^2} + 2 \imath \frac{m}{\hbar}
\frac{\partial \Psi_{mn}^S}{\partial t} = 0,
\end{equation}
and the non-relativistic form of the Bateman-Hillion ansatz
\begin{equation} \label{eq: schrodinger_solution}
\Psi_{mn}^{OS} = \Phi_{mn}^S(\xi_1, \xi_2, \tau) \exp
\left[\frac{\imath}{\hbar}(P_3 x_3 -E_s t) \right],
\end{equation}
where $E_S$ is the non-relativistic energy of the particle and
\begin{equation} \label{eq: bateman_schrodinger_connection}
\Phi_{mn}^S = \int \Phi_{mn} \delta(\xi_3-v\tau) d\xi_3.
\end{equation}
For comparison to results in the present context $\Psi_{mn}^{OS}$
must be further subject to the unitary transformation (\ref{eq:
unitary_transformation}) that simplifies to
\begin{equation} \label{eq: unitary_transformation_nr}
\Psi_{mn}^S = \Psi_{mn}^{OS} \exp \left( \frac{\imath N^{mn}\hbar t}{m_0w_0^2} \right)
\end{equation}
in the the non-relativistic limit $c \rightarrow \infty$.

It is readily shown that eq. (\ref{eq: continuity_Schrodinger}) is
the continuity equation for the Schr\"{o}dinger equation (\ref{eq:
schrodinger_wave_equation}) since
\begin{equation}
\frac{\partial j_3}{\partial x_3} = \frac{P_s}{\hbar}
\frac{\partial}{\partial x_3} |\Psi_{mn}^S|^2 = 0.
\end{equation}
It is thus concluded from a direct comparison of eqs. (\ref{eq: continuity_BH_time_form}) and ({\ref{eq: continuity_Schrodinger}) that
\begin{equation}
P_{BH} = m_0 \frac{j_3 + j_4}{k_3+k_4} = |\Psi_{mn}|^2
\end{equation}
is the relativistic probability density for finding a particle in a
Bateman-Hillion beam. This differs from the widely cited \cite{RPF}
Klein-Gordon probability density
\begin{equation}
P_{KG} = \frac{j_4}{c}
\end{equation}
due to the fact $\Psi_{mn}$ is further constrained under the
parabolic equation (\ref{eq: KG_bateman}). It is also of interest to
notice that $P_{BH}$ is form invariant under Lorentz transformations
whereas $P_{KG}$ is not as an isolated component of a 4-vector.

Bateman-Hillion functions can be normalized using the integral expression
\begin{equation} \label{eq: normalization}
\int_{-\infty}^{+\infty} \int_{-\infty}^{+\infty}
\int_{-\infty}^{+\infty} |\Psi|^2 \delta(\xi_3-v_3\tau)d\xi_1 d\xi_2 d \tau  = \frac{1}{L},
\end{equation}
having set the probability of finding the particle in a beam of
length $L$ is 1. This evaluates to
\begin{equation}
C_{mn}^{HG} = \sqrt{\frac{ 2}{\pi w_0^2 L 2^{m+n}m!n!}}
\end{equation}
for Hermite-Gaussian beams; and
\begin{equation}
C_{lp}^{LG} = \sqrt{\frac{ 4p!}{w_0^2 L (p+|l|)!}}
\end{equation}
for Laguerre-Gaussian beams.

Expectation values for the measurable properties of each particle in
the beam can be calculated as
\begin{multline} \label{eq: expectations}
\langle \Psi |\hat{O}| \Psi \rangle_P = \\
\int_{-\infty}^{+\infty} \int_{-\infty}^{+\infty}
\int_{-\infty}^{+\infty} (\Psi^* \hat{O} \Psi)
\delta(\xi_3-v_3\tau)d\xi_1 d\xi_2 d \tau,
\end{multline}
where $\hat{O}$ is the quantum mechanical operator for each
observable quantity. Here, the subscript $P$ has been included as a
reminder that the integration is performed over a planar
cross-section perpendicular to the axis of the beam but not along
the axis itself.

\section{IV. Calculation of 4-Momentum}
The canonical 4-momentum operator $\hat{p}_\mu$ is defined in eq.
(\ref{eq: canonical_momentum_operators})
in terms of the 4-position vector $x_\mu$. We next seek to use the fact
$\Psi_{mn}$ depends on $X_{\mu}$ as well as $x_{\mu}$ to define a distinct
kinetic 4-momentum operator $\hat{P}_{\mu}$ to satisfy the eigenvalue
equation
\begin{equation} \label{eq: kinetic_momentum_def}
\hat{P}_\mu \Psi_{mn}=\hbar k_\mu \Psi_{mn}
\end{equation}
The first step is to write
\begin{multline}
\Phi_{mn}(\xi_1, \xi_2, \xi_3+c\tau) = \\
\Phi_{mn}(x_1-X_1, x_2-X_2, x_3-X_3+ct-cT)
\end{multline}
having used eq. (\ref{eq: relative_coordinates}). This indicates
\begin{equation}
\frac{\partial \Phi_{mn}}{\partial x_\mu} = -\frac{\partial \Phi_{mn}}{\partial X_\mu}
\end{equation}
and therefore
\begin{equation}
\imath \hbar \left( \frac{\partial}{\partial x^\mu} + \frac{\partial}{\partial X^\mu} \right) \Psi_{mn} = \hbar\left(k_\mu + \kappa_\mu^{mn} \right)\Psi_{mn}
\end{equation}
From comparison of this expression to eq. (\ref{eq: kinetic_momentum_def})
it can be seen that 
\begin{equation} \label{eq: kinetic_4_momentum}
\hat{P}_\mu \Psi_{mn} = \left( \imath\hbar \frac{\partial}{\partial x^\mu} + \imath\hbar \frac{\partial}{\partial X^\mu} - \hbar \kappa_\mu^{mn} \right) \Psi_{mn} = \hbar k_\mu \Psi_{mn}
\end{equation}
or equivalently
\begin{equation} \label{eq: momentum_eigenvalue_equations}
\hat{P}_1 \Psi_{mn}=\hat{P}_2 \Psi_{mn} = 0, \quad \hat{P}_3
\Psi_{mn} = \hbar k_3 \Psi_{mn},
\end{equation}
\begin{equation} \label{eq: energy_eigenvalue_equations}
\hat{P}_4 \Psi_{mn}= \sqrt{\hbar^{2}
k_3^2+2 \kappa^{mn} \left( k_3 + k_4 \right)+m_0^2c^2} \Psi_{mn},
\end{equation}
having used eq. (\ref{eq: dispersion_relation}). These results are the eigenvalue
equations for the kinetic 4-momentum of each particle in a relativistic Hermite-Gaussian
beam. In completing this argument, it is necessary to find the explicit form of
$N^{mn}$ from eq. (\ref{eq: four_momentum_expected}).

Inserting the Bateman-Hillion ansatz (\ref{eq: bateman_solution})
into eq. (\ref{eq: four_momentum_expected}) gives
\begin{equation} \label{eq: expectected_momentum}
\langle \Psi_{mn}|m_0\hat{j}_\mu| \Psi_{mn} \rangle_P=\hbar (k_\mu + \kappa_\mu^{mn}) + \langle \Phi_{mn}
|m_0\hat{j}_\mu| \Phi_{mn} \rangle_P.
\end{equation}
Here, the term $\langle \Phi_{mn}|m_0\hat{j}_\mu| \Phi_{mn} \rangle_P$ can
be evaluated using the integrals
\begin{equation} \label{eq: gaussian_integtals_1}
\int_{-\infty}^{+\infty} x H_m^2(\sqrt{\alpha} x)
e^{-\alpha x^2}dx = 0,
\end{equation}
\begin{equation} \label{eq: gaussian_integtals_2}
\int_{-\infty}^{+\infty} x^2 H_m^2(\sqrt{\alpha} x)
e^{-\alpha x^2}dx = \sqrt{\frac{\pi}{\alpha^3}}\left(\frac{1}{2}+m \right).
\end{equation}
The result is
\begin{equation} \label{eq: current_kappa_connection}
\langle \Phi_{mn}|m_0\hat{j}_\mu| \Phi_{mn} \rangle_P = -\hbar \kappa^{mn}
\end{equation}
having set
\begin{equation} \label{eq: N_def}
N^{mn} = 1 + m + n.
\end{equation}
Putting eq. (\ref{eq: current_kappa_connection}) into (\ref{eq: expectected_momentum}) gives
\begin{equation}
\langle \Psi_{mn}|m_0\hat{j}_\mu| \Psi_{mn} \rangle_P = \hbar k_\mu
\end{equation}
It is thus established that the eigenvalues of the kinetic 4-momentum
operator $\hat{P}_\mu$ are equal to the expectations values for the mass current
for all Hermite-Gaussian beam modes.

Equations (\ref{eq: energy_eigenvalue_equations}) and (\ref{eq: N_def}) 
enable the total energy $E_{HG}^{mn}$ for each particle in a Hermite-Gaussian 
mode to be written as
\begin{equation} \label{eq: energy_HG_modes}
E_{HG}^{mn}= c\sqrt{\hbar^{2} k_3^2+ \frac{2 \hbar^{2}}{w_0^2}(1+m+n)+m_0^2c^2}.
\end{equation}
Comparing this result to the energy of a free particle
\begin{equation} \label{eq: energy_free_mass}
E_{FP}= c\sqrt{\hbar^{2} k_3^2 + m_0^2c^2}
\end{equation}
of identical mass $m_0$ and axial wave number $k_3$ shows that the beam 
particle picks up an additional energy contribution 
\begin{equation} \label{eq: middle_term}
\hbar^2 K_T^{mn} = \frac{2 \hbar^{2}}{w_0^2}(1+m+n)
\end{equation}
where $K_T^{mn}$ is defined in eq. (\ref{eq: mystery_term}), as a 
result of being localized. The remaining task is therefore is to assign 
a physical interpretation to this term.

It can be inferred from inspection of eq. (\ref{eq: particle_current_def}) 
that the expectation values of canonical 4-momentum and mass current must 
be related through the expression
\begin{equation} \label{eq: momentum_current_relationship}
\langle \Psi_{mn}|m_0\hat{j}_\mu| \Psi_{mn} \rangle_P = \Re \langle \Psi_{mn}|\hat{p}_\mu| \Psi_{mn} \rangle_P
\end{equation}
where the operator $\Re$ takes the real part of the argument.
Equations (\ref{eq: operator_relationship_3}), (\ref{eq: expectected_momentum}), 
(\ref{eq: current_kappa_connection}) and (\ref{eq: momentum_current_relationship}) 
can therefore be used together to give
\begin{equation}
\Re \langle \Psi_{mn} |\hat{p}_1^2 + \hat{p}_2^2| \Psi_{mn} \rangle_P =
\frac{2\hbar^{2}}{w_0^2}(1 + m + n)
\end{equation}
This shows that the middle term under the square root sign in 
eq. (\ref{eq: energy_HG_modes}}) represents the contribution of the 
fluctuating transverse components of momentum to the total energy of 
each particle. 

\section{V. Quantum Potential}
The concept of distinguishing between canonical and kinetic
4-momentum has familiarity from the description \cite{RPF} of a
particle of charge $e$ moving in an electromagnetic 4-potential
$A_\mu$. The kinetic 4-momentum for this problem is
\begin{equation} \label{eq: em_kinetic}
\hat{\pi}_\mu = \hat{p}_\mu - eA_\mu.
\end{equation}
For the purposes of comparison the relationship between the kinetic
and the canonical 4-momentum of a beam particle given in eq.
(\ref{eq: kinetic_4_momentum}) can be written as
\begin{equation} \label{eq: qm_kinetic}
\hat{P}_\mu = \hat{p}_\mu - m_0\hat{U}_\mu,
\end{equation}
where
\begin{equation} \label{eq: qm_potential_def}
\hat{U}_\mu =\frac{\hbar}{m_0} \left( \frac{1}{\imath}
\frac{\partial}{\partial X^\mu} + \kappa_\mu^{mn} \right).
\end{equation}
Eqs. (\ref{eq: em_kinetic}) and (\ref{eq: qm_kinetic}) are similar
in form but $A_\mu$ is an external 4-potential whereas $\hat{U}_\mu$
is an operator. The understanding here is that wave
equations are constructed using kinetic 4-momentum to take account of
external potentials and canonical 4-momentum if no external potential
is present. The Hermite-Gaussian function $\Psi_{mn}$ was derived from
a wave equation that contains only canonical 4-momentum operators but it is
still possible to identify a 4-potential like term $\hat{U}_\mu$ in the
definition of the kinetic 4-momentum $P_\mu$ analogous to the role
of the external 4-potential $A_\mu$ in $\pi_\mu$. Equation (\ref{eq:
qm_potential_def}) will be referred to as the 4-potential operator.

Kinetic 4-momentum was defined in eq. (\ref{eq: kinetic_momentum_def}) to
be a real quantity. It follows from eq. (\ref{eq: qm_kinetic}) that the particle
current can be written in the form
\begin{equation} \label{eq: qm_kinetic_real}
j_\mu =  \left( \frac{\hbar k_\mu}{m_0}  + U_\mu \right) |\Psi|^2
\end{equation}
where
\begin{equation}
U_\mu = \frac{\hbar}{m_0} \Re \left( \frac{\imath}{\Psi}
\frac{\partial \Psi}{\partial X^\mu} + \kappa_\mu^{mn} \right)
\end{equation}
is a real quantum 4-potential field. Comparing eq. (\ref{eq: qm_kinetic_real})
to the component eqs. (\ref{eq: paricle_current_1}) through (\ref{eq: paricle_current_4})
gives
\begin{equation} \label{eq: qp_1}
U_1^{mn} = \frac{\hbar}{m_0} \frac{4b(\xi_3+c \tau) \xi_1}{w_0^2[(\xi_3+c \tau)^2 +4b^2]},
\end{equation}
\begin{equation} \label{eq: qp_2}
U_2^{mn} = \frac{\hbar}{m_0} \frac{4b(\xi_3+c \tau) \xi_2}{w_0^2[(\xi_3+c \tau)^2 +4b^2]},
\end{equation}
\begin{multline} \label{eq: qp_3}
U_3^{mn} = \frac{\hbar}{m_0}  \left[ \kappa^{mn}
- \frac{2b(1+m+n)}{(\xi_3+c \tau)^2+4b^2} \right] \\
- \frac{\hbar}{m_0} \frac{2b(\xi_1^2+\xi_2^2)[(\xi_3+c \tau)^2-4b^2]}{w_0^2[(\xi_3+c \tau)^2+4b^2]^2} ,
\end{multline}
\begin{multline} \label{eq: qp_4}
U_4^{mn} = \frac{\hbar}{m_0} \left[ - \kappa^{mn} + \frac{2b(1+m+n)}{(\xi_3+c \tau)^2+4b^2} \right]  \\
+ \frac{\hbar}{m_0} \frac{2b(\xi_1^2+\xi_2^2)[(\xi_3+c \tau)^2-4b^2]}{w_0^2[(\xi_3+c \tau)^2+4b^2]^2},
\end{multline}
to be the explicit form of the quantum 4-potential for a Hermite-Gaussian beam.

Expression (\ref{eq: qm_kinetic_real}) is a quantum mechanical equation describing a particle in a
localized state. In the absence of localization $(w_0 \rightarrow \infty)$
it reduces to the form
\begin{equation} \label{eq: qm_kinetic_free}
j_\mu = \frac{\hbar k_\mu}{m_0} |\Psi|^2
\end{equation}
showing that the quantum 4-potential term has vanished. Squaring eq. (\ref{eq: qm_kinetic_free})
gives
\begin{equation} \label{eq: qm_kinetic_free_squared}
j_M^2 - m_0^2c^2 |\Psi|^2 = 0
\end{equation}
where $j_M^2=m_0^2j_\mu j^\mu$. It is clear that if we now add back the quantum 
4-potential into both eqs. (\ref{eq: qm_kinetic_free}) and 
(\ref{eq: qm_kinetic_free_squared}) then eq. (\ref{eq: qm_kinetic_free_squared})
must pick up an additional scalar term $V^2$ such that
\begin{equation} \label{eq: qm_kinetic_squared_1}
j_M^2 - m_0^2c^2 |\Psi|^2 + V^2 = 0
\end{equation}
where
\begin{equation} \label{eq: qm_kinetic_squared_2}
V^2 = -|\hbar k_\mu - m_0U_\mu|^2 - m_0^2c^2 |\Psi|^2
\end{equation}
Expanding this expression gives 
\begin{multline} \label{eq: rel_scalar_potential}
V^2 = -m_0^2|U_\mu^{mn}|^2 - 2 \hbar k^\mu m_0 U_\mu^{mn} - \hbar^2 K_T^{mn} = \\
4\hbar^2 \left[\frac{1+m+n}{w^2} - \frac{\xi_1^2+\xi_2^2}{w^4}\right]
\end{multline}
having used
\begin{equation}
|U_\mu^{mn}|^2 = -\frac{\hbar^2}{m_0^2} \frac{16b^2(\xi_3+c \tau)^2 (\xi_1^2+\xi_2^2)}{w_0^4[(\xi_3+c \tau)^2 +4b^2]^2},
\end{equation}
\begin{multline}
k^\mu U_\mu^{mn} = \frac{\hbar}{m_0} \left[ - \frac{K_T^{mn}}{2} + \frac{8b^2(1+m+n)}{w_0^2[(\xi_3+c \tau)^2+4b^2]} \right]  \\
+ \frac{\hbar}{m_0} \frac{8b^2(\xi_1^2+\xi_2^2)[(\xi_3+c \tau)^2-4b^2]}{w_0^4[(\xi_3+c \tau)^2+4b^2]^2},
\end{multline}
alongside eq. (\ref{eq: dispersion_relation}). It is concluded from this 
argument that $V^2$ is itself a quantum potential appearing in eq. 
(\ref{eq: qm_kinetic_squared_2}) as the scalar analog of the quantum 
4-potential $U_\mu$ in eq. (\ref{eq: qm_kinetic_real}).

The OAM operator (\ref{eq: OAM_operator}) can be rewritten in Cartesian
coordinates to give
\begin{equation}
\hat{L}_{3} = \xi_1 \hat{p}_2 - \xi_2 \hat{p}_1
\end{equation}
or equivalently
\begin{equation}
\hat{L}_{3} = \xi_1 (\hat{P}_2 + m_0\hat{U}_2) - \xi_2 (\hat{P}_1 + m_0\hat{U}_1)
\end{equation}
having used eq. (\ref{eq: qm_kinetic}). This last result simplifies to
\begin{equation}
\hat{L}_{3} = \xi_1 m_0\hat{U}_2 - \xi_2 m_0\hat{U}_1
\end{equation}
since $P_\mu = (0,0,\hbar k_3, \hbar k_4)$. It is therefore
concluded that the quantum 4-potential operator $\hat{U}_\mu$ and
not the kinetic 4-momentum operator $\hat{P}_\mu$ is the source of
the mass flow resulting in OAM.

Calculating the expectation value of each component $\hat{U}_{\mu}$ of
the quantum 4-potential and the scalar analog $V^2$ we obtain
\begin{equation}
\langle \Psi_{mn}|U_{\mu}| \Psi_{mn} \rangle_P = \langle \Psi_{mn}|V_{\mu}^2| \Psi_{mn} \rangle_P = 0,
\end{equation}
This result shows that quantum 4-potential is a fluctuating phenomenon. Specifically, the
presence of quantum 4-potential can cause the canonical 4-momentum of a localized particle
in a beam to instantaneously deviate from the kinetic 4-momentum but it has no affect at all
on the expected 4-momentum of the particle.

The original concept of a quantum potential was introduced by David
Bohm \cite{Bohm} who started from an ansatz to solve the
Schr\"{o}dinger equation. This takes the form
\begin{equation} \label{eq: bohm_ansatz}
\Psi = R\exp\left(\imath \frac{S}{\hbar} \right),
\end{equation}
where the amplitude $R$ and  $S / \hbar$ are real valued
functions. 

On inserting eq. (\ref{eq: bohm_ansatz}) into the Schr\"{o}dinger
equation (\ref{eq: schrodinger_wave_equation}), the imaginary part of the
equation can be identified as the continuity equation (\ref{eq:
continuity_Schrodinger}) and the real part as the Hamilton-Jacobi
equation
\begin{equation} \label{eq: hamilton_jacobi}
-\frac{\partial S_{mn}}{\partial t} = \frac{|\nabla S_{mn}|^2}{2m_0}
+ Q.
\end{equation}
where
\begin{equation} \label{eq: bohm_potential}
Q = -\frac{\hbar^2}{2m_0} \frac{\nabla^2 R_{mn}}{R_{mn}},
\end{equation}
is the Bohm potential. It is of interest next to investigate how the 
quantum 4-potential and the Bohm potential are related to each other.

The solution to the Schr\"{o}dinger equation for Hermite-Gaussian
beams is given in eqs. (\ref{eq: schrodinger_solution}) and
(\ref{eq: bateman_schrodinger_connection}). On comparing eq.
(\ref{eq: schrodinger_solution}) and (\ref{eq: bohm_ansatz}) the
explicit form of the amplitude $R_{mn}$ and phase function $S_{mn}$
can be read off to be
\begin{multline} \label{eq: schrodinger_amplitude}
R_{mn} = \frac{C_{mn}^{HG} w_0}{w_S}H_m\left(
\frac{\sqrt{2}\xi_1}{w_S}\right) H_n\left(
\frac{\sqrt{2}\xi_2}{w_S}\right)\\ \times \exp \left( -\frac{ \xi_1^2+\xi_2^2}{w_S^2} \right),
\end{multline}
and
\begin{multline} \label{eq: schrodinger_phase}
S_{mn} = P_3 x_3 - Et - (1+m+n)\hbar\omega_0 t \\
+ \frac{2 (\xi_1^2+\xi_2^2)\hbar \omega_0\tau}{w_S^2 }
- \hbar(1+m+n)\arctan \left( 2 \omega_0 \tau \right),
\end{multline}
where
\begin{equation}
w_S = w_0 \sqrt{1+4 \omega_0 \tau^2}, \quad \omega_0 =
\frac{\hbar}{m_0 w_0^2}.
\end{equation}
Inserting eq. (\ref{eq: schrodinger_amplitude}) into eq. (\ref{eq: bohm_potential})
shows the Bohm quantum potential for a non-relativistic Hermite-Gaussian beam to be
\begin{equation} \label{eq: bohm_beam_potential}
Q = \frac{2\hbar^2}{m_0} \left[\frac{1+m+n}{w_S^2} + \frac{\xi_1^2+\xi_2^2}{w_S^4} \right].
\end{equation}
Eq. (\ref{eq: bohm_beam_potential}) can in turn be inserted into the
Hamilton-Jacobi equation (\ref{eq: hamilton_jacobi}) giving
eq. (\ref{eq: schrodinger_phase}) as a solution.

It is clear from eqs. (\ref{eq: rel_scalar_potential}) and (\ref{eq: bohm_beam_potential}) that
\begin{equation}
Q = \lim_{c \rightarrow \infty} \frac{V^2}{2m_0}.
\end{equation}
This result shows that the Bohm potential for a Hermite-Gaussian
beam is the non-relativistic limit of the scalar form $V^2$ of the
relativistic quantum potential defined in eq. (\ref{eq: rel_scalar_potential}).

\section{VI. Summary}

A relativistic solution for Hermite-Gaussian particle beams
presented in an earlier paper \cite{RD2} has been used to calculate
the properties of the particles in the beam. In the original paper,
the solutions were obtained using a Bateman-Hillion ansatz that
reduces the Klein-Gordon equation to a parabolic form thus enabling
|$\Psi|^2$ to be interpreted as the probability density for finding
the particle. It was shown the solutions are form preserving under
Lorentz transformations and correspond to those of the
Schr\"{o}dinger equation in the non-relativistic limit. It was also
shown the solutions take account of the Gouy phase in the beam.

In this paper, a Lorentz covariant kinetic 4-momentum operator has
been introduced equal to canonical 4-momentum operator minus a
quantum 4-potential term. The quantum 4-potential originates at the
beam waist where it introduces fluctuating terms into the canonical 
4-momentum of transversely localized particles. All the eigenvalues 
of the kinetic 4-momentum operator have in fact been shown to equal 
the expectation values of the real parts of the canonical 4-momentum 
components. The total energy of a particle for each beam mode has
also been calculated. It has been found, in particular, that the energy
of a particle in a beam differs from the energy of a free particle as
a result of fluctuating transverse momentum components in the spatial
plane perpendicular to the axis of the beam.

Transverse momentum is needed to explain both the divergence of the
beam after passing through the beam waist as well as OAM. Here,
solutions have been presented for Laguerre-Gaussian modes to demonstrate
the possibility for OAM in the Bateman-Hillion formalism. It has also
been found that in our proposed partitioning of canonical 4-momentum
into kinetic and quantum 4-potential parts that the kinetic part
makes no contribution to OAM meaning that OAM is a pure manifestation
of quantum 4-potential. A clear indicator to this is that particles
must be localized to exhibit OAM. Free particles cannot have OAM in the
absence of localization since they have no quantum 4-potential.

The quantum 4-potential has been discussed in relation to the
electromagnetic 4-potential and the Bohm potential. Quantum
4-potential acts on mass in an analogous manner to how electromagnetic
4-potential operates on charge. Specifically, both potentials operate to
produce a distinction  between the canonical 4-momentum of a particle
that includes the influence of the 4-potential and a kinetic 4-momentum
that does not. It is clear though that the quantum and electromagnetic
4-potentials are, at least, different in the sense that a charged particle
can intrinsically generate an electromagnetic field whereas it is the
localization of a particle that indicates quantum 4-potential and not just 
the presence of the particle by itself.

The Bohm potential and quantum 4-potential related concepts that both
vanish in the absence of localization and have null expectation
values. The quantum 4-potential has been developed here in the context of 
relativistic Hamiltonian quantum mechanics. By contrast the Bohm potential 
was first identified as a term in a quantum form of the non-relativistic
Hamilton-Jacobi equation that Bohm derived from the real part of the
Schr\"{o}dinger equation. It has been demonstrated here that if both sides
of a quantum mechanical equation containing quantum 4-potential are squared
then the quantum potential in the derived equation is a scalar term. The 
Bohm potential is simply the non-relativistic limit of this scalar counterpart
of quantum 4-potential.

\end{document}